\documentclass{article}
\usepackage{spconf,amsmath,graphicx}

\usepackage{enumitem}
\setlist{nosep, leftmargin=14pt}

\usepackage{mwe} 
\usepackage{multirow}
\usepackage[utf8]{inputenc}
\usepackage{animate}
\usepackage{graphicx}
\usepackage{amssymb} 
\usepackage{color}
\usepackage{hyperref}

\title{JoB-VS: Joint Brain-Vessel Segmentation in TOF-MRA Images
}
\name{Natalia Valderrama$^{\star}$, Ioannis Pitsiorlas $^{\dagger}$, Luisa Vargas$^{\star}$, Pablo Arbeláez$^{\star}$, Maria A. Zuluaga $^{\dagger}$ }

\address{$^{\star}$ Center of Formation and Research in Artificial Intelligence, Universidad de los Andes, Colombia \\ 
$^{\dagger}$ Data Science Department, EURECOM, Sophia Antipolis, France}

\begin{document}
%
\maketitle
\begin{abstract} 
We propose the first joint-task learning framework for brain and vessel segmentation (JoB-VS) from Time-of-Flight Magnetic Resonance images. Unlike state-of-the-art vessel segmentation methods, our approach avoids the pre-processing step of implementing a model to extract the brain from the volumetric input data. Skipping this additional step makes our method an end-to-end vessel segmentation framework. JoB-VS uses a lattice architecture that favors the segmentation of structures of different scales (e.g., the brain and vessels). Its segmentation head allows the simultaneous prediction of the brain and vessel mask. Moreover, we generate data augmentation with adversarial examples, which our results demonstrate to enhance the performance. JoB-VS achieves 70.03\% mean AP and 69.09\% F1-score in the OASIS-3 dataset and is capable of generalizing the segmentation in the IXI dataset. These results show the adequacy of JoB-VS for the challenging task of vessel segmentation in complete TOF-MRA images.


\end{abstract}
\section{Introduction} 
\label{sec:intro}


3D brain vessel tree segmentation is critical for diagnosing, managing, treating, and intervening in a wide range of conditions with large population-level implications~\cite{captcha}. Despite being a widely studied problem~\cite{MOCCIA201871} and the surge of  automated deep learning approaches to address it, existing pipelines require a pre-processing step to generate a brain mask to remove non-brain tissue signal from the images~\cite{captcha,livne2019u,ds6}. As a result, it is not possible to achieve cerebrovascular segmentation in an end-to-end fashion but through a two-step approach where brain masks are first extracted and then fed to the brain vessel segmentation algorithm.

\begin{figure}[htb]
  \centering
\includegraphics[width=0.85\linewidth]{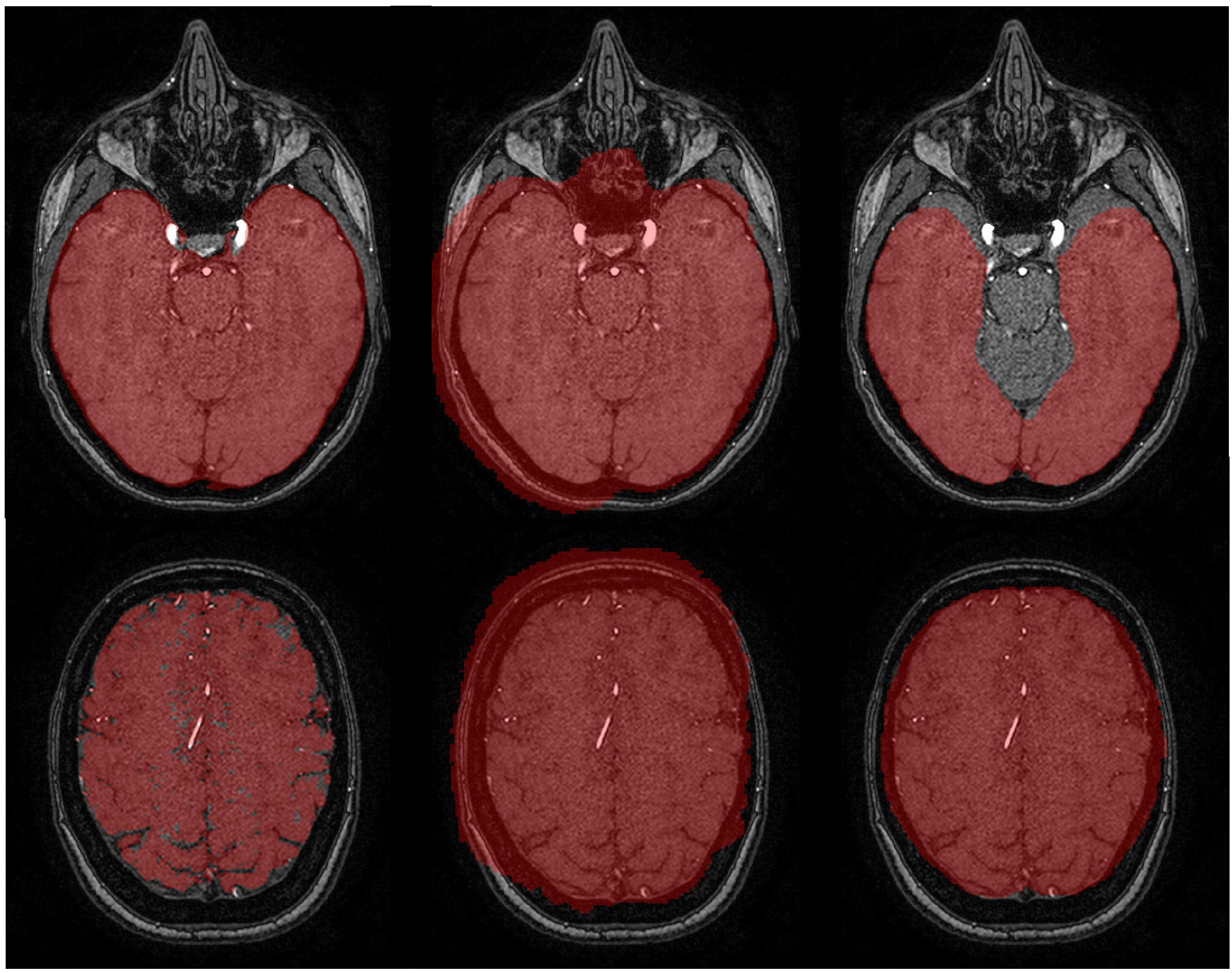}
  \caption{TOF-MRA brain masks obtained with SynthStrip~\cite{HOOPES2022119474} (left), ROBEX~\cite{iglesias2011} (center) and HD-BET~\cite{isensee2019} (right). These methods are not designed nor tested for neurovascular images.}
  \label{fig:brain_masks}
\end{figure}

Although there are several well-established algorithms to generate brain masks (see Sec. 2 in~\cite{HOOPES2022119474}), the need for such a step within a cerebrovascular tree segmentation pipeline poses two problems. On the one hand, it represents an additional computational burden, as an additional model needs to be trained for the sole purpose of brain extraction (also known as skull stripping). On the other hand, although most brain extraction techniques generalize well across multiple brain image modalities~\cite{HOOPES2022119474,iglesias2011,isensee2019}, these methods have not been designed or tested for vessel-specific brain image modalities, such as Time of Flight (TOF) Magnetic Resonance Angiography (MRA). This condition leads to poor-quality results that need to be manually corrected before the brain mask can be used as input to the vessel segmentation step (Fig.~\ref{fig:brain_masks}).
In this work, we propose a Joint Brain-Vessel Segmentation framework, JoB-VS, A joint-task learning approach that enables simultaneous brain and brain vessel segmentation. JoB-VS achieves a true end-to-end segmentation by avoiding the additional step of generating a brain mask for every unseen image to be segmented. We evaluate the performance of the proposed JoB-VS framework in a cohort of TOF-MRA images from two well-known datasets, OASIS-3~\cite{LaMontagne2019.12.13.19014902} and IXI~\footnote{\url{https://brain-development.org/}}, demonstrating our method's capacity to achieve competitive results for both vessel and brain segmentation.
To ensure the reproducibility of our results, we make our code publicly available at \url{https://github.com/BCV-Uniandes/JoB-VS}.

\begin{figure*}[htb]
  \centering
  \includegraphics[width=1\linewidth]{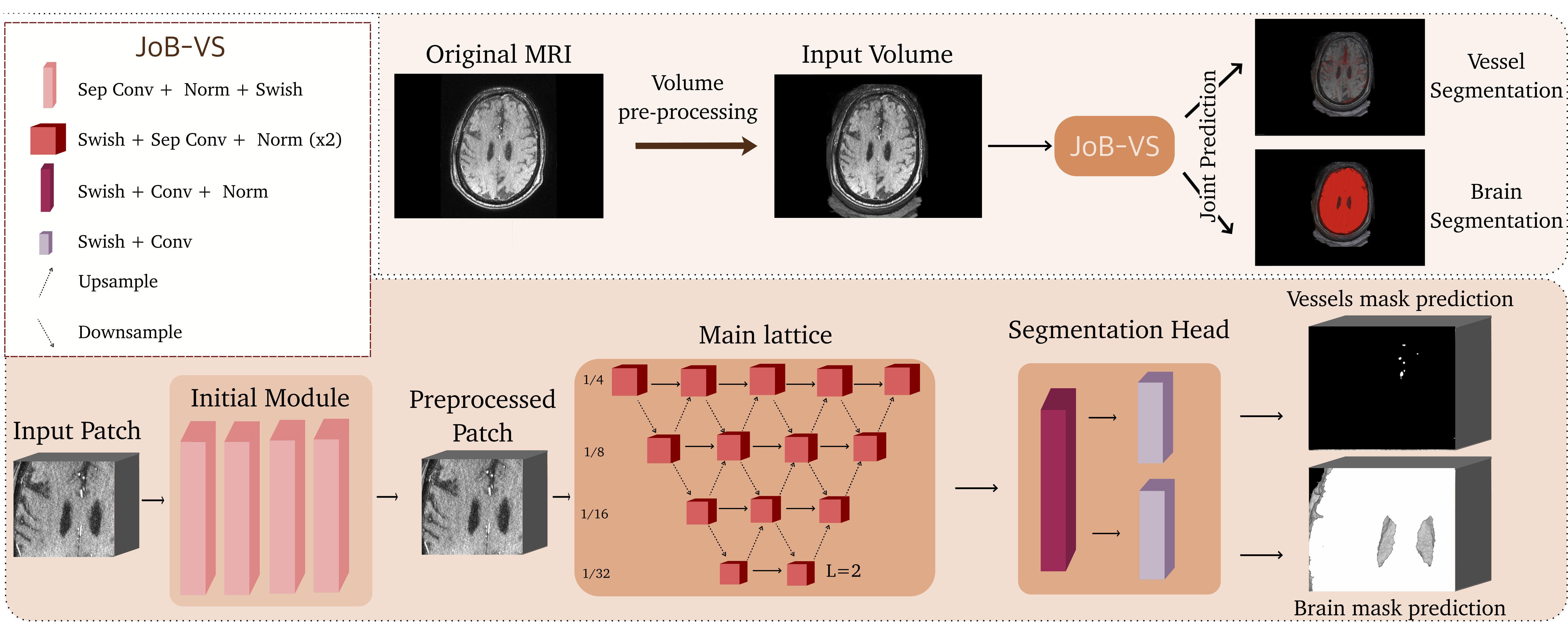}
 
  \caption{\textbf{Overview of JoB-VS}. Our end-to-end approach enables a simultaneous brain and vessel segmentation mask prediction using complete TOF-MRA images. Our method builds upon ROG \cite{rog}, a lattice 3D segmentation method, which we modify with two segmentation heads to perform independent brain and vessel masks. Best viewed in color. }
 \label{fig:overview}
\end{figure*}

\section{Method} 
\label{sec:method}




Figure \ref{fig:overview} depicts our Joint Brain-Vessel Segmentation (JoB-VS) framework. Differently from previous works \cite{captcha,livne2019u,ds6}, JoB-VS avoids a separate skull-stripping step to remove the non-brain signal from the images at training. Instead, we jointly train a 3D segmentation model to simultaneously predict separate label masks for both vessels and the brain tissue. 

We build upon the RObust Generic medical image segmentation framework (ROG)~\cite{rog}. ROG performs segmentation in volumetric data leveraging large receptive fields and high-resolution features thanks to its three-stage architecture consisting of an initial module, a triangular-shaped lattice, and a segmentation head. The lattice arrangement, which connects every node with upper and lower-level nodes, preserves the advantages of multi-scale processing, which is crucial for vessel segmentation. Additionally, by controlling the lattice length, $L$, ROG can favor different types of anatomical structures. By setting $L=2$, we manage to achieve a better perception of small structures, such as the blood vessels, and at the same time, at the lower levels of the lattice, it preserves information about larger structures, like the brain. 

We modify the segmentation head to cope with multiple tasks rather than the original single-task formulation. The JoB-VS segmentation head has two convolutions and upsampling factors, yielding two outputs: the vessel and brain masks. We consider equally the predictions of the vessels, $y'_{v}$, and the brain mask, ${y'_{b}}$, when estimating the total loss during the joint learning process:

\begin{equation}
    \mathcal{L}_{total} = \alpha\mathcal{L}_{brain}(y'_{b}, y_{b}) + \beta\mathcal{L}_{vessel}(y'_{v}, y_{v}) \label{eq:loss}
\end{equation}
with ${y_{v}}$ and ${y_{b}}$ being the vessel and the brain ground truth annotations, respectively. The weights $\alpha$ and $\beta$ control the contribution of each loss to the total loss. According to our experimental results, we set both parameters to 1. For each of the losses, $\mathcal{L}_{vessel}$ and $\mathcal{L}_{brain}$, we use a combination of the Dice loss and the Cross-Entropy, as described in ~\cite{nn_unet}: 
\begin{equation}
    \mathcal{L}_{vessel} = \mathcal{L}_{brain} = \mathcal{L}_{dice} + \mathcal{L}_{CE} 
    \label{eq:loss_di_ce}
\end{equation}

\noindent
\textbf{Free Adversarial Training Fine-tuning (AT FT).}
A common challenge of deep learning methods for vessel segmentation is the difficulty of obtaining annotated data due to the complexity of vascular trees and the small size of vessels~\cite{captcha}. To compensate for the small size of the annotated set, we fine-tune JoB-VS using generated adversarial data augmentation. For this purpose, we rely on ``Free'' Adversarial Training (AT)~\cite{free_at}, which generates gradient-based perturbations on the input data during training, allowing us to simultaneously update the model weights and the input perturbations over N iterations on the same mini-batch. The strength of the perturbation is controlled by the parameter $\epsilon$, which we set in $8/255$, and we set N=5.


\noindent
\textbf{Implementation details.} 
The implementation of the deep learning models was done using PyTorch. 
We train and test the JoB-VS framework with a batch size of 1 in a single NVIDIA TITAN Xp GPU. We use Adam with a weight decay of $1\mathrm{e}{-5}$, an initial learning rate of $5\mathrm{e}{-4}$, and the scheduler described in~\cite{rog}. Generally, we train during 1000 epochs; nonetheless, the training epochs may vary as we use an early stop system when the learning rate is too low. We use the data augmentation, input patch selection criteria, and inference process used in \cite{rog}.

\begin{table*}[htb]
\centering
\caption{Method comparison. We report average mean Average Precision (mAP) and average max F1-score in the BM and NBM test setup with standard deviation within our two-fold cross-validation and for the two training setups (BM and NBM).  The best and second-best results are shown in bold and underlined, respectively.}
\label{tab:main_results}
\resizebox{1\textwidth}{!}{%
\begin{tabular}{ll|rr|rr|rr}
\hline
\multicolumn{1}{c}{\multirow{2}{*}{\begin{tabular}[c]{@{}c@{}c@{}}Training\\ setup\end{tabular}}} &
  \multicolumn{1}{c|}{\multirow{2}{*}{Model}} &
  \multicolumn{2}{c|}{mean AP (\%)} &
  \multicolumn{2}{c|}{F1-score (\%)} &
  \multicolumn{2}{c}{clDice (\%)}\\
\multicolumn{1}{c}{} &
  \multicolumn{1}{c|}{} &
  \multicolumn{1}{c}{BM} & 
  \multicolumn{1}{c|}{NBM} & 
   \multicolumn{1}{c}{BM} & 
  \multicolumn{1}{c|}{NBM} & 
  \multicolumn{1}{c}{BM} & 
  \multicolumn{1}{c}{\begin{tabular}[c]{@{}c@{}}NBM\end{tabular}} \\ \hline
\multirow{3}{*}{\begin{tabular}[c]{@{}l@{}}BM\end{tabular}} &
  Half U-net \cite{livne2019u} &  68.98 $\pm$ 3.26 & 23.96 $\pm$ 0.05  & \underline{77.05 $\pm$ 0.06} &  36.84 $\pm$ 0.54 & \underline{78.02 $\pm$ 0.64} &  34.35 $\pm$ 1.90 \\
 &	
  Residual U-Net \cite{residual_unet}& 77.02 $\pm$ 1.77 & 41.33 $\pm$ 9.76 & 74.35 $\pm$ 1.92 & 49.95 $\pm$ 6.63 &  75.10 $\pm$ 2.59 &  39.12 $\pm$ 5.86 \\
 &
  Single-VS & \textbf{81.13 $\pm$ 2.01} & 51.05 $\pm$ 9.78 & \textbf{78.64 $\pm$ 1.88} & 57.48 $\pm$ 5.99 & \textbf{80.22 $\pm$ 1.22} &  60.24 $\pm$ 3.90 \\ \hline
\multirow{4}{*}{\begin{tabular}[c]{@{}l@{}}NBM\end{tabular}} &
  Half U-net \cite{livne2019u} & 10.71 $\pm$ 14.80 & 26.91 $\pm$ 37.71 & 19.51 $\pm$ 26.90 & 33.71 $\pm$ 46.99 & 12.49 $\pm$ 17.67 & 35.48 $\pm$ 50.18 \\
 &
  Residual U-Net \cite{residual_unet} & 73.43 $\pm$ 0.52 & 48.49 $\pm$ 8.43 & 70.77 $\pm$ 0.55 & 54.85 $\pm$ 4.77 & 69.52 $\pm$ 3.55 & 49.87 $\pm$ 2.97
   \\
 &
  Single-VS &
  74.64 $\pm$ 1.62 & \underline{66.08 $\pm$ 5.75} & 74.87 $\pm$ 1.15 & \underline{67.21 $\pm$ 3.31} & 74.63 $\pm$ 4.02 &  \underline{73.72 $\pm$ 1.72} \\ \hline
 NBM & JoB-VS (Ours) &
  \underline{77.69 $\pm$ 1.63} & \textbf{70.03 $\pm$ 4.31} & 74.98 $\pm$ 0.58 & \textbf{69.09 $\pm$ 3.31} & 77.24 $\pm$ 1.56  & \textbf{74.56 $\pm$ 1.04}
   \\ \hline
\end{tabular} }
\end{table*}

\section{Experiments and Results} 
\label{sec:exps}
\subsection{Experimental Setup} 
\paragraph*{Datasets and Setup.} 
\label{sec:data}
We use a cohort of high-resolution 57 and 572 TOF-MRA images from the OASIS-3~\cite{LaMontagne2019.12.13.19014902} and the Information eXtraction from Images (IXI) datasets, respectively. TOF-MRA from OASIS-3 have volume dimensions 576 $\times$ 768 $\times$ 232 and voxel size 0.3$\times$0.3$\times$0.6 mm$^3$. All of them have been previously annotated (i.e., vessel and brain labels) by an experienced rater and a neurologist~\cite{captcha}. Images from IXI have an average volume dimension of 576 $\times$ 768 $\times$ 160 and an average voxel size 0.47$\times$0.47$\times$0.80 mm$^3$. No labels are available for the dataset.

All images were pre-processed following the guidelines in \cite{rog,nn_unet}. We re-sample the volumes with the median voxel spacing in the TOF-MRAs to tackle heterogeneous patient scanning. Furthermore, we apply z-score normalization for each image in the dataset and perform further intensity modification with global statistics. In particular, we clip the intensities to the [0.5, 99.5] percentiles of the vessel values and perform z-score normalization using the mean and standard deviation of all intensity values in the dataset.

We use two-fold cross-validation to train and quantitatively assess our method with the OASIS-3 TOF-MRA images. The folds contain images for 29 and 28 subjects. As there are no available publicly labels for the IXI dataset, we use this cohort of 572 TOF-MRA scans for qualitative assessment of the generalization capabilities of our framework. 
\\
\noindent
\textbf{External baselines.} We compare the proposed JoB-VS framework to three vessel segmentation approaches. Two of them are state-of-the art techniques for 3D brain vessel segmentation, Half U-net~\cite{livne2019u} and DS6~\cite{ds6}, whereas the third one has been used for retinal vessel extraction~\cite{residual_unet}. Additionally, we train a single-task vessel segmentation method (Single-VS) adopting the original version of ROG for general medical image segmentation~\cite{rog}.  

\noindent
\textbf{Evaluation metrics.} 
We evaluate the vessel segmentation task with 
the mean Average Precision (mAP), the cldice metric \cite{shit2021cldice}, and the maximum F1-score (F1), which is equivalent to the Dice Similarity coefficient (DSC). 


We justify this use of detection task metrics in vessel segmentation, considering that tubular structures in three dimensions are analogous to linear structures in two dimensions. We evaluate the performance of the brain segmentation task using the DSC. For the metrics of both tasks, we report the mean and standard deviation between the fold's performance.


\begin{figure}[htb]
  \centering
  \includegraphics[width=1\linewidth]{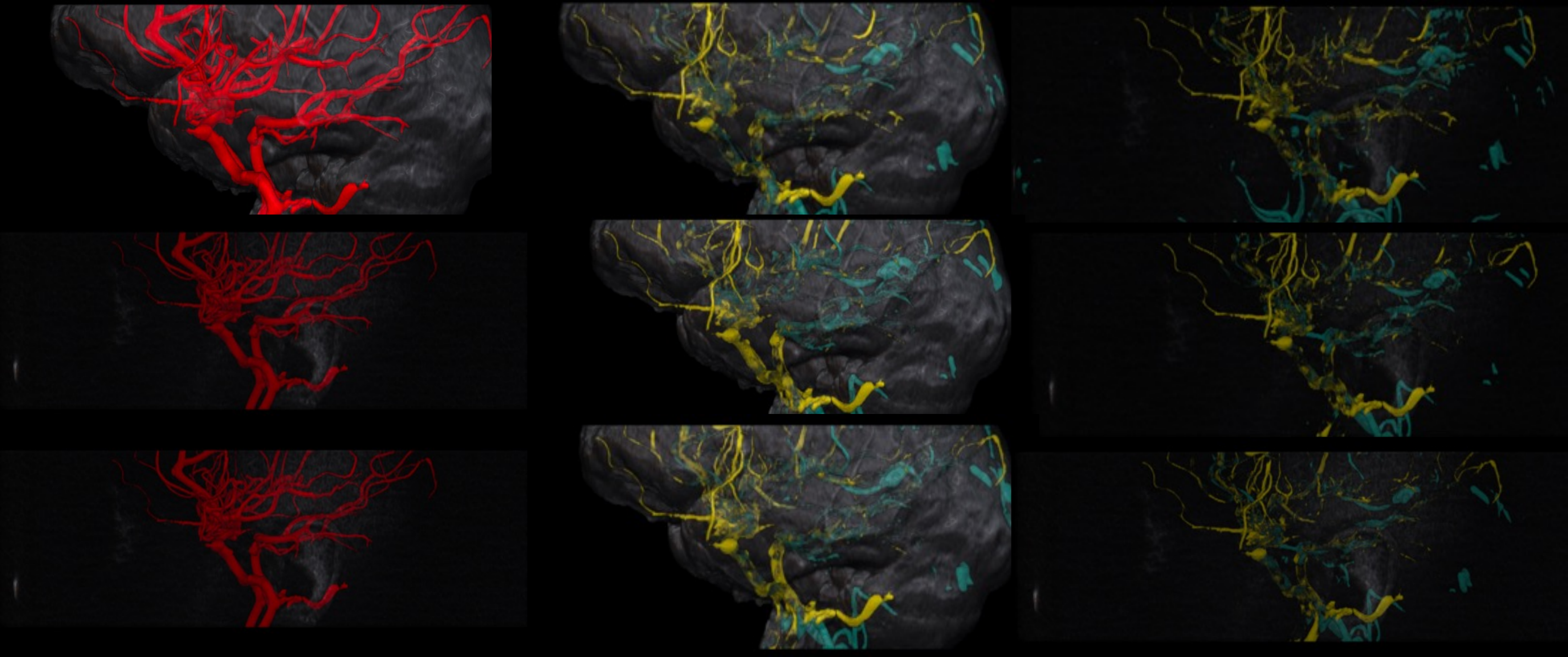}
  \caption{3D renderings of vessel segmentation results on an OASIS-3 subject. From left to right: TOF-MRA and ground truth, BM segmentation, and NBM segmentation. From top to bottom: Single-VS BM, Single-VS NBM, and JoB-VS NBM. Note that in the Single-VS BM model (first row), the TOF-MRA input was masked by the brain. Green indicates false positive, and yellow indicates false negative.}
  \label{fig:qualitative_results}
\end{figure}
\subsection{Method Comparison} 
\label{sec:comparison}
Table~\ref{tab:main_results} summarizes the performance of the trained models with the different setups. For a fair comparison with the external baselines, we use two experimental setups for training and testing. First, we consider the standard approach used by most vessel segmentation techniques, where a brain mask (BM) is required during training to consider only the image signal within the brain and discard any potential vessel annotations outside the brain mask region. Second, we train the methods using the original TOF-MRA images, with no brain mask (NBM) required. In both scenarios, we evaluate the results obtained in the test set with BM (ground truths for baselines, predicted mask for JoB-VS) and without them (NBM). We omit results from DS6, as we could not reach a performance similar to that reported in the original publication~\cite{ds6}.


Overall, all methods achieve over 68\% performance in all metrics when tested with a brain mask (BM), independently of whether they were trained with or without a brain mask. However, all baselines report a drop in performance when trained in the NBM setup (i.e., 58.27\%, 3.59\%, 6.49\%, in mAP, for Half U-Net, Residual U-Net, and Single-VS). When we test the models without using a brain mask at inference (NBM), the evaluation metrics significantly drop, i.e., 45.02\%, 24.94\%, 8.56\% in mAP, for Half U-Net, Residual U-Net, and Single-VS. The results expose their high dependence on brain masks at training and testing to achieve good results.


JoB-VS results suggest that joint training may help the model learn to delimit the relevant anatomical structures (the vessels) within the brain. This outcome is reflected in its highest performance (mAP=70.03\%, F1=69.09\%, clDice=74.56\%) in the NBM test setup. We further illustrate this behavior in Figure \ref{fig:qualitative_results}, where models trained with BM struggle to differentiate vessels from anatomical structures with similar intensities outside the brain (i.e., the skull), indicating they may be learning a threshold and not leveraging other characteristics of vessel anatomy. By providing information about the whole image (i.e., no masking), JoB-VS avoids this problem. 

Despite the positive effects of avoiding masking at training, the results suggest that using a brain mask at test time is always beneficial, even for JoB-VS, which improves its performance (mAP= 77.69\%, F1=74.98\%, clDice=77.24\%). In this respect, the joint segmentation of brain and vessel labels is advantageous, as the former can be directly used to ameliorate the latter without needing any additional method.

\begin{table}[]
\caption{Joint segmentation vs. single-task training before and after using Adversarial Training fine-tuning (AT FT). We report DSC for brain and mAP for vessel segmentation.}
\label{tab:ablation}
{%
\begin{tabular}{ll|cc}
\hline
Model type &
\multicolumn{1}{c|}{\begin{tabular}[c]{@{}c@{}}AT FT \end{tabular}} &
 \begin{tabular}[c]{@{}c@{}}Brain\\ DSC (\%)\end{tabular} & \begin{tabular}[c]{@{}c@{}}Vessel\\ mAP (\%)\end{tabular} \\ \hline
Single-task & x & 94.19 $\pm$ 0.22 & 63.09 $\pm$ 7.58 \\
Single-task & \checkmark & 96.29 $\pm$ 0.08 & 66.67 $\pm$ 7.61 \\ \hline
JoB-VS & x & 95.60 $\pm$ 1.00 & 68.72 $\pm$ 5.92 \\ 
JoB-VS & \checkmark & 95.73 $\pm$ 0.74 & \textbf{70.03 $\pm$ 4.31} \\ \hline
\end{tabular}
}
\end{table}

\subsection{Ablation Experiments} 
%

We study the advantages of the joint-task framework and adversarial training fine-tuning (AT FT). To this end, we train single-task models of ROG \cite{rog} for 1) brain segmentation and 2) vessel segmentation with input volumes where the brain was extracted using the predictions of the single-task brain model. We test the single-task vessel model in the BM setup using the brain mask predictions from the trained brain model. We also assess the performance JoB-VS method before and after AT FT in the NBM test setup. Table \ref{tab:ablation} summarizes these models' performance, evidencing the benefit of AT FT as it increases the performance by 1.31\% and 3.58\% in mAP in the single and joint-task models. 

Moreover, the brain segmentation model's performance is critical to vessel segmentation with single-task methods. Our brain segmentation model has a DSC of 96.29\%,  which results in 66.67\% mAP in the vessel segmentation. However, a method with 100\% DSC (predictions same as ground truth) would result in vessel segmentation with 81.13\% mAP (see Sec. \ref{sec:comparison}). Instead, the joint model improves the vessel segmentation task without dependency on a brain mask generator and its performance.

\subsection{IXI Dataset and Monai APP} 

\begin{figure}[t!]
  \centering
  \includegraphics[width=1\linewidth]{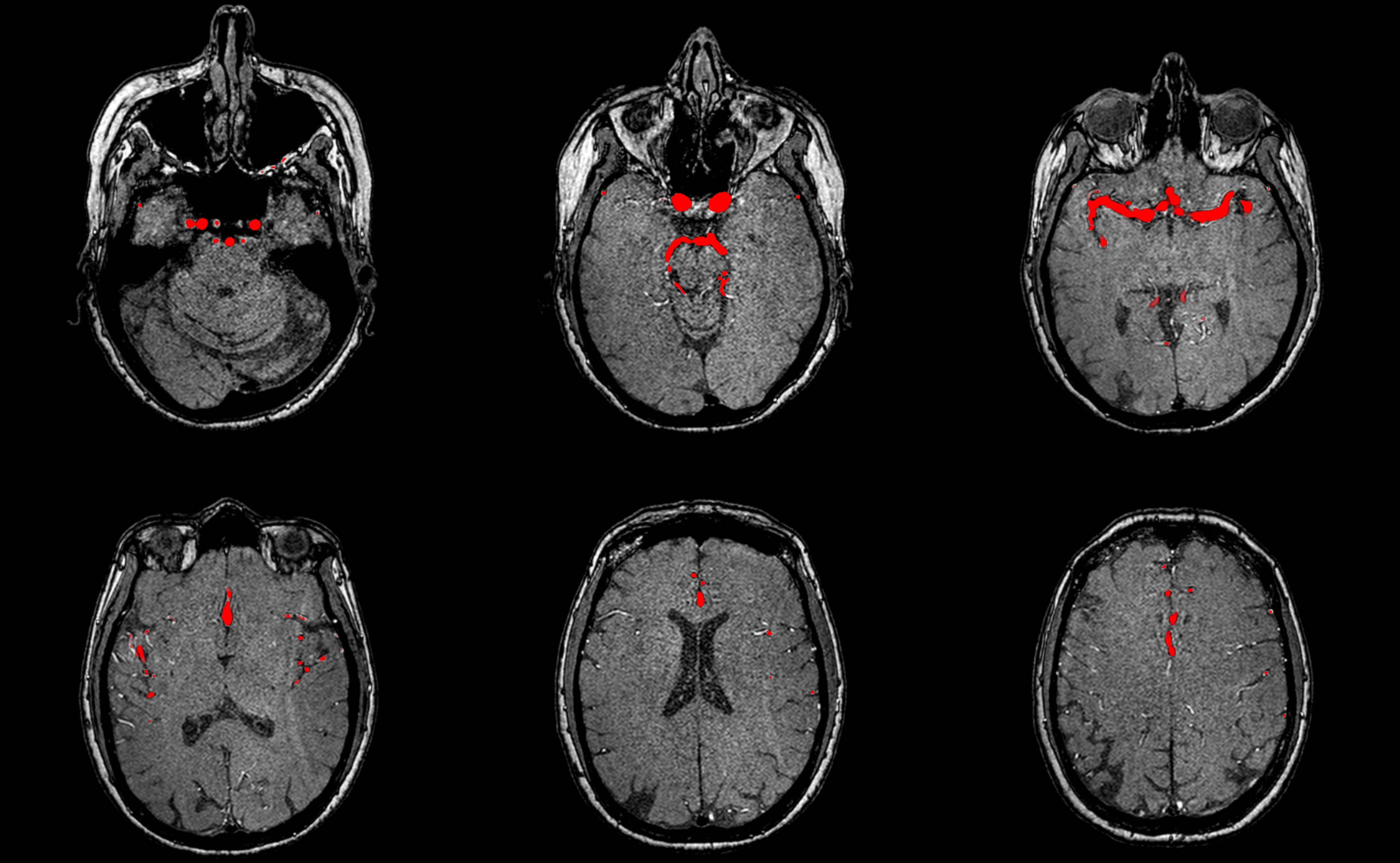}
  \caption{Vessel segmentation results on IXI (subject IXI605-HH-2598). The vessel mask is shown in red.}
  \label{fig:ixi}
\end{figure}

We test the generality power of our best model by making inferences in the IXI Dataset. Figure \ref{fig:ixi} presents an example illustrating how our joint-task approach can correctly identify and differentiate vessels from other anatomical structures. Additionally, we develop a MONAI~\cite{monailabel} app 
to load our method and model weights. We will make this app publicly available to facilitate vessel annotation and the interactive refinement of JoB-VS' predicted vessel masks.




\section{Conclusions} 

We introduce JoB-VS, a joint-task learning method for brain and vessel segmentation in TOF-MRA images. 
By avoiding the additional step of skull stripping, we propose an end-to-end vessel and brain segmentation framework. In addition, we employ ``Free'' Adversarial Training to compensate for the limited training data size, typical in 3D brain vessel segmentation. Our results demonstrate the benefits of joint training and AT fine-tuning while proving competitive performance in the vessel segmentation task across the OASIS-3 and IXI datasets. Lastly, we develop a tool to contribute to further expanding the limited datasets in this challenging task.

\newpage
\paragraph*{Compliance with ethical standards.}
This research study was conducted retrospectively using human subject data made available in open access by the OASIS-3~\cite{LaMontagne2019.12.13.19014902} and the IXI datasets. Ethical approval was not required, as confirmed by the license attached to the open-access databases.

\paragraph*{Acknowledgments.}
\label{sec:acknowledgments}
NV thanks the support of UniAndes-DeepMind Scholarship 2021. MAZ is supported by the French government via the 3IA Côte d’Azur Investments in the Future project managed by the ANR  (ANR-19-P3IA-0002) and by the ANR JCJC project I-VESSEG (22-CE45-0015-01). The authors declare no conflicts of interest.

\bibliographystyle{IEEEbib}
\bibliography{strings,refs}
\end{document}